\documentclass[journal,transmag]{IEEEtran}

\usepackage{hyperref,color,breqn,multirow}
\usepackage{amssymb}
\usepackage{amsmath}
\usepackage{mathtools}
\usepackage{siunitx}
\usepackage{subfig}

\newcommand{\Npod}{N_{\mathrm{POD}}}
\newcommand{\Ninit}{N_{\mathrm{init}}}
\newcommand{\Ntrain}{N_{\mathrm{train}}}

\usepackage{eso-pic}
\AddToShipoutPicture*{\footnotesize\sffamily\raisebox{0.75cm}{\hspace{1.65cm}\fbox{\parbox{\textwidth}{\copyright 2023 IEEE.  Personal use of this material is permitted.  Permission from IEEE must be obtained for all other uses, in any current or future media, including reprinting/republishing this material for advertising or promotional purposes, creating new collective works, for resale or redistribution to servers or lists, or reuse of any copyrighted component of this work in other works.}}}}

\begin{document}

\title{Reduced Basis Approximation for Maxwell's Eigenvalue Problem \\and Parameter-Dependent Domains}

\author{\IEEEauthorblockN{Max Kappesser\IEEEauthorrefmark{1},
Anna Ziegler\IEEEauthorrefmark{1}, and
Sebastian Schöps\IEEEauthorrefmark{1}, 
}
\IEEEauthorblockA{\IEEEauthorrefmark{1}Computational Electromagnetics Group, Technische Universität Darmstadt, Darmstadt, Germany}%
\thanks{Manuscript received xxxx, 2023; revised xxxx. 
Corresponding author: A.~Ziegler (email: anna.ziegler@tu-darmstadt.de).}}

\markboth{Journal of \LaTeX\ Class Files,~Vol.~14, No.~8, August~2015}%
{Shell \MakeLowercase{\textit{et al.}}: Bare Demo of IEEEtran.cls for IEEE Transactions on Magnetics Journals}

\IEEEtitleabstractindextext{%
\begin{abstract}
In many high-frequency simulation workflows, eigenvalue tracking along a parameter variation is necessary. This can become computationally prohibitive when repeated time-consuming eigenvalue problems must be solved. Therefore, we employ a reduced basis approximation to bring down the computational costs. It is based on the greedy strategy from Horger et al. 2017 which considers multiple eigenvalues for elliptic eigenvalue problems. We extend this algorithm to deal with parameter-dependent domains and the Maxwell eigenvalue problem. 
In this setting, the reduced basis may contain spurious eigenmodes, which require special treatment.
We demonstrate our algorithm in an eigenvalue tracking application for an eigenmode classification.
\end{abstract}

}

\maketitle

\IEEEdisplaynontitleabstractindextext

\IEEEpeerreviewmaketitle

\begin{figure}[!t]
\centering
\def\figheight{0.07\textheight}
\subfloat[]{\includegraphics[trim = 350 300 300 280, clip, height = \figheight]{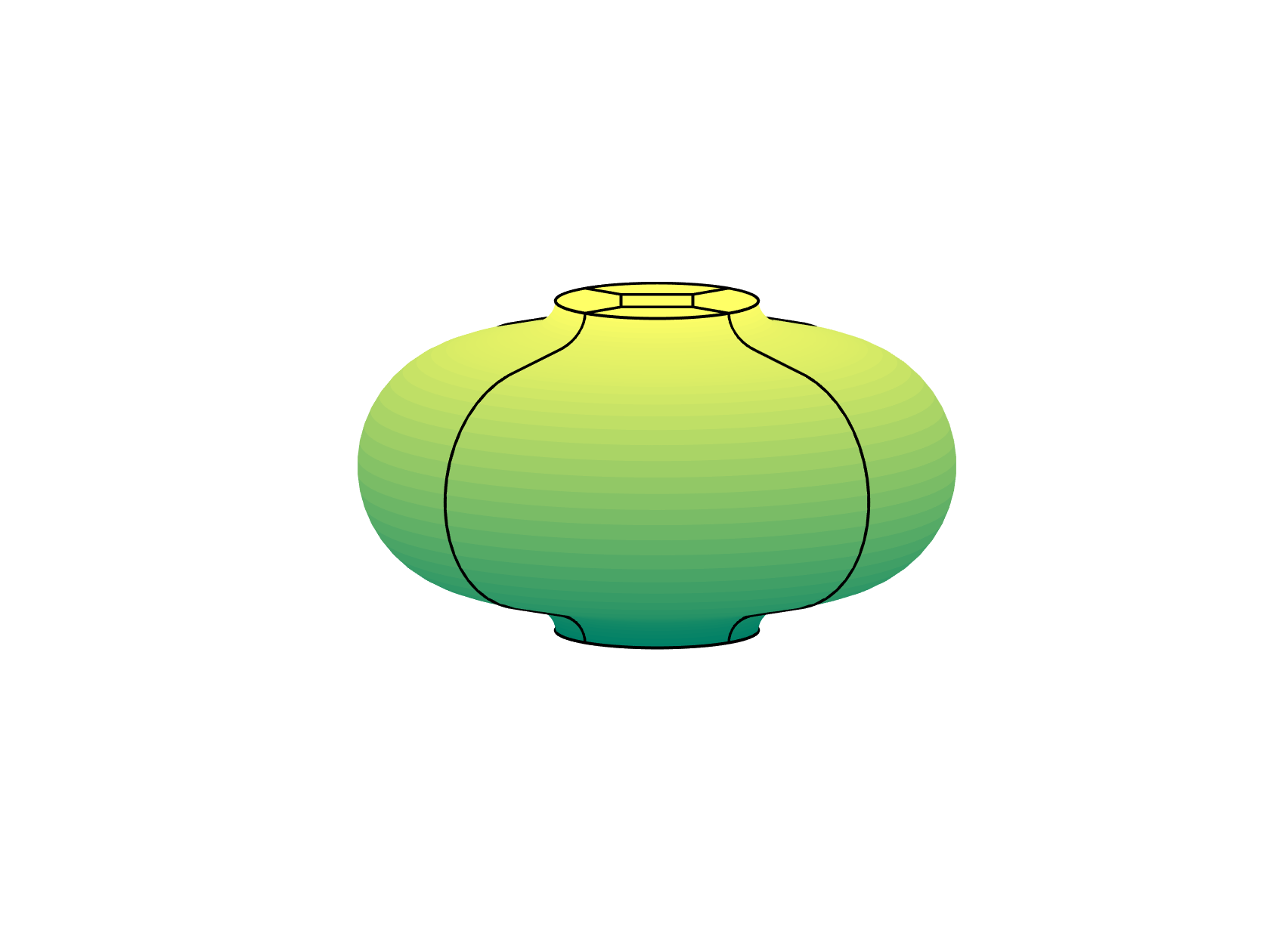}%
\label{fig:t10}}
\subfloat[]{\includegraphics[trim = 350 300 300 280, clip, height = \figheight]{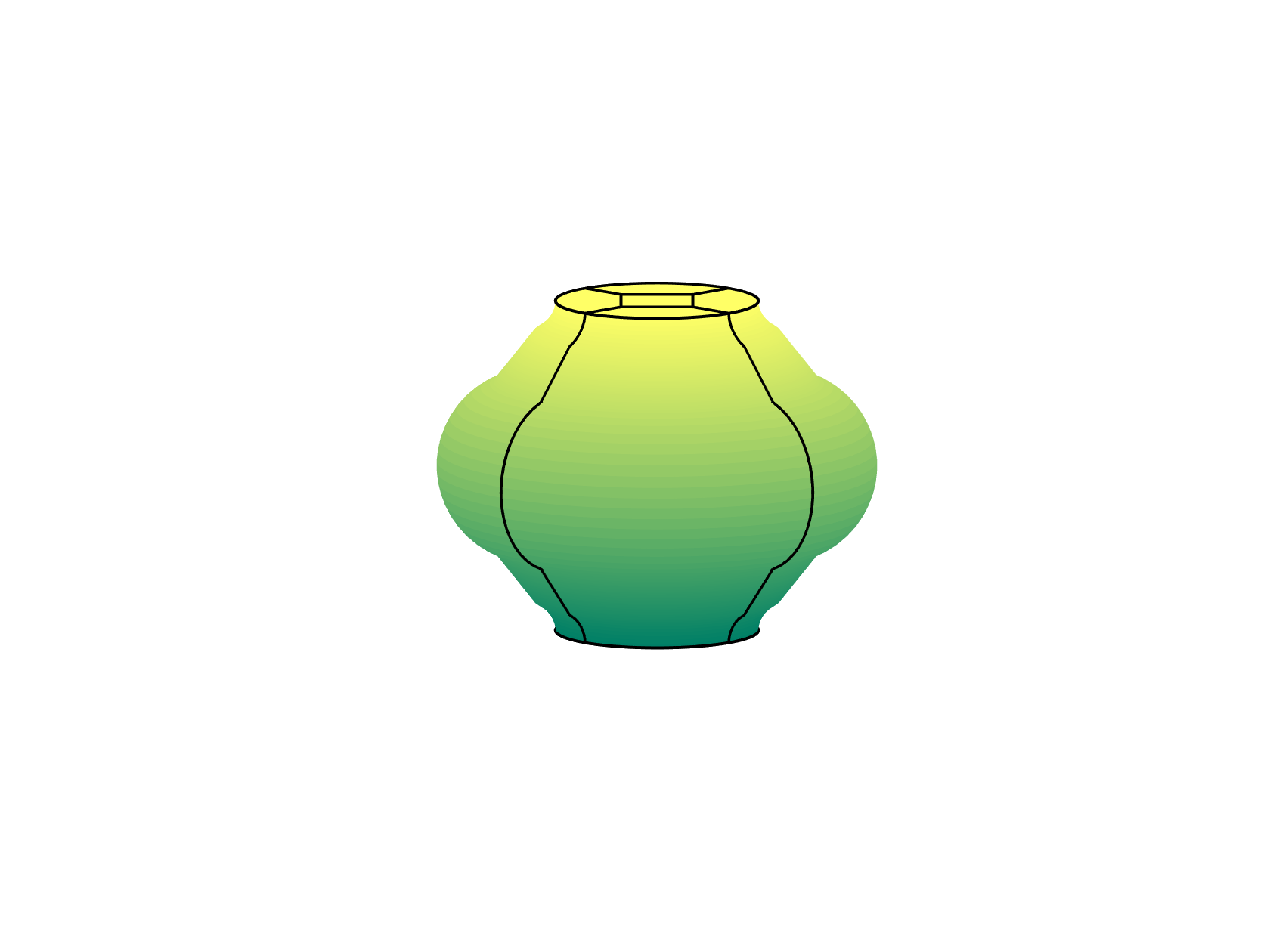}}%
\label{fig:t06}\\
\subfloat[]{\includegraphics[trim = 350 300 300 280, clip, height = \figheight]{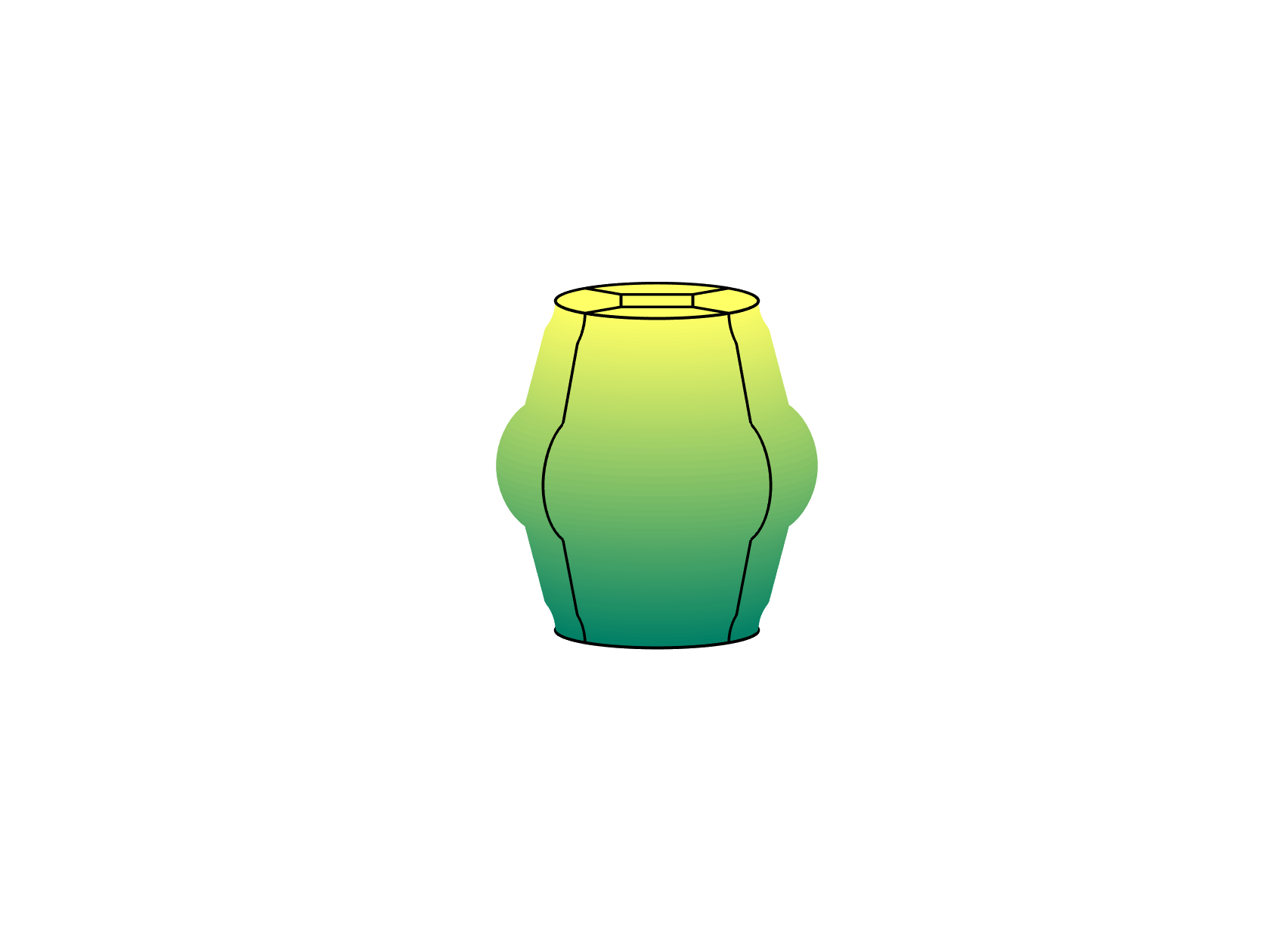}%
\label{fig:t03}}
\subfloat[]{\includegraphics[trim = 350 300 300 280, clip, height = \figheight]{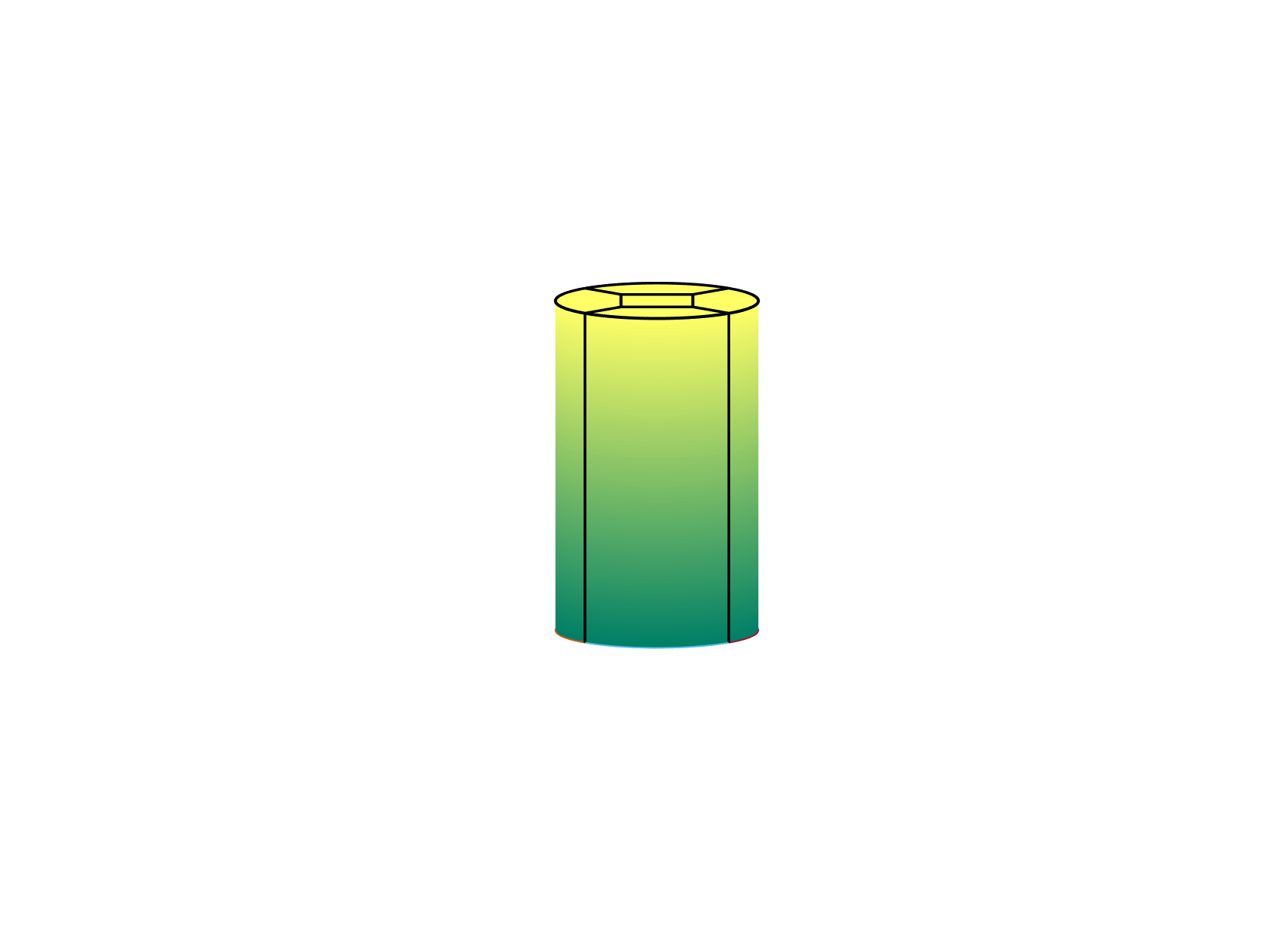}}%
\label{fig:t00}
\caption{Morphing of the 1-cell TESLA cavity (a) to the pillbox cavity (d). The black lines indicate the patch boundaries when discretizing the domain with splines.}
\label{fig:morphing}
\end{figure}

\IEEEPARstart{E}{igenvalue tracking} is used in various engineering applications. It requires repeated solutions of similar eigenvalue problems in order to be able to identify a specific eigenmode along a parameter variation, e.g. in an optimization setting. To this end, various methods have been proposed, which gradually follow the variation, extrapolate the modes and use a matching criterion to restore the correct order in the case of a crossing, see e.g.  \cite{Lui_1997aa,Jorkowski_2018aa}.  
In this paper, we employ the eigenvalue tracking procedure to classify the eigenmodes of a complex-shaped superconducting electromagnetic resonator, i.e. the TESLA cavity \cite{Aune_2000aa}. 
This requires morphing the TESLA geometry to a pillbox cavity, as depicted in \autoref{fig:morphing}, and following the eigenvalues along the deformation.
The full procedure is described in~\cite{Ziegler_2023aa}. 

In the following, we show how the tracking procedure can be sped up using model order reduction. 
Our approach is inspired by \cite{Horger_2017aa} which deals with eigenvalue problems from solid mechanics with parameter-dependent materials. 
Therefore, the idea is extended to Maxwell's eigenvalue problem on parameterized domains.
We remove spurious modes from the basis and subsequently apply it within the tracking procedure. 

In \autoref{sec:problem}, we introduce the problem statement and give an overview of the eigenvalue tracking procedure.
We recall the proper orthogonal decomposition and the greedy strategy for the construction of the reduced basis in \autoref{sec:rb} and describe the treatment of the spurious modes in \autoref{sec:spuriousModes}.
Finally, we demonstrate our work with numerical results in \autoref{sec:results} and conclude the work in \autoref{sec:conclusion}.

\section{Problem Statement}\label{sec:problem}

For the electromagnetic cavity, we consider the Maxwell eigenproblem on the bounded, simply connected domain $\Omega\in\mathbb{R}^3$ with Lipschitz boundary $\partial \Omega$. We use the source-free time-harmonic formulation in vacuum
\begin{equation}
\nabla\times\left(\nabla\times\mathbf{E}\right)={\omega^{2}}/{c_{0}^2}\mathbf{E}  \quad\text{in }\Omega
\label{eq:maxwellProblem}
\end{equation}
with $\mathbf{E}\times\mathbf{n}=0$ on $\partial \Omega$, where $\mathbf E$ denotes the electric field strength, $\mathbf{n}$ the outwards pointing normal vector and $c_0$ the speed of light in vacuum. 
In order to be able to numerically solve problem~\eqref{eq:maxwellProblem}, we discretize it using Isogeometric Analysis (IGA), where we use B-splines and NURBS as basis functions for analysis and geometry~\cite{Vazquez_2010aa}.
We introduce the deformation parameter~$t \in T = [0,1]$ for parametrizing the domain $\Omega(t)$ and its boundary.
Using IGA for the discretization allows for a smooth and straightforward domain deformation while maintaining suitable mesh quality without remeshing. 
After spatial discretization, we obtain the generalized eigenvalue problem
\begin{equation} 
\mathbf{A}(t)\mathbf{v}(t) = \lambda(t) \mathbf{B}(t) \mathbf{v}(t)
\label{eq:eigproblem}
\end{equation} 
with parameter-dependent eigenvector~$\mathbf{v}(t)$ and eigenvalue~$\lambda(t)$. 
The $\mathcal{N}$-dimensional, parameter-dependent stiffness matrix~$\mathbf{A}$ is expressed by
\begin{equation}
    \mathbf{A}_{i,j}(t) = \int_{\Omega(t)}\textrm{curl}\,\mathbf{w}_{j}\cdot\textrm{curl}\,\mathbf{w}_{i}\,\mathrm{d}\mathbf{x}
\end{equation}
and mass matrix~$\mathbf{B}$ computed via 
\begin{equation}
    \mathbf{B}_{i,j}(t) = \int_{\Omega(t)}\mathbf{w}_{j} \cdot\mathbf{w}_{i} \,\mathrm{d}\mathbf{x},
\end{equation}
where all basis functions~$\mathbf{w}_i$ and~$\mathbf{w}_j$ with $i,j=1,\ldots,\mathcal{N}$, are chosen from a finite-dimensional subspace of~$H_0(\mathrm{curl};\Omega(t))$, the space of the square-integrable vector fields with square-integrable curl and vanishing trace. 
We can then compute the discrete eigenpairs $\left(\mathbf{v}({t}), \lambda({t})\right)$ and the eigen\-frequency $f(t)={\sqrt{\lambda(t)}}c_0/{(2\pi )}$. 
Details on the discretization and the consideration of the parameter dependence of the domain can be found in~\cite{Ziegler_2023aa}.

\subsection{Eigenvalue Tracking}
\begin{figure}[!t]
\includegraphics[]{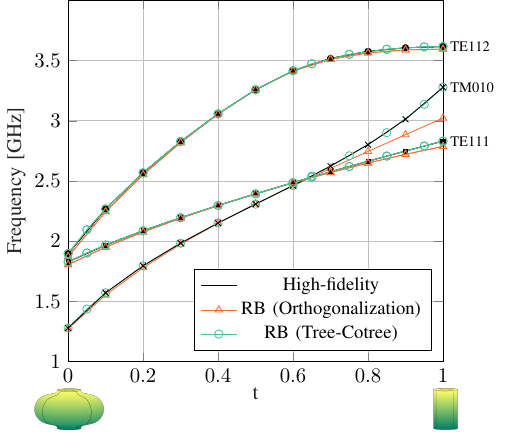}
\caption{Tracking of the first five eigenvalues from the TESLA cavity ($t=0$) to the pillbox ($t=1$) and matching with the analytical solutions and the corresponding classification. The black lines correspond to the solutions on the high-fidelity system, the orange lines to the solutions on the reduced basis (RB), where the spurious modes were removed by Gram-Schmidt orthogonalization, see \autoref{sec:gramschmidt}. The green lines indicate the solutions on the reduced basis, where the spurious modes were removed with the tree-cotree gauge, see \autoref{sec:treecotree}.}
\label{fig:tracking}
\end{figure}

We denote the domain of the TESLA cavity by $\Omega(t=0)$ and of the pillbox cavity by $\Omega(t=1)$.
For the classification of the eigenmodes of the TESLA cavity, we solve the problem
\begin{equation}
\begin{bmatrix}
			\mathbf{A}(t)\mathbf{v}(t)-\lambda(t) \mathbf{B}(t)\mathbf{v}({t}) \\
			\mathbf{c}^{\mathrm{H}}\mathbf{B}(t)\mathbf{v}(t)-1 
		\end{bmatrix}
		=\mathbf{0}
\end{equation}		
with a suitable vector $\mathbf{c}$ for normalization. 
We always track~$K$ eigenvalues at the same time, possibly even more than we need at the beginning, since the eigenvalues may cross along the deformation in a way that is unknown in advance, c.f. \autoref{fig:tracking}.

By solving
\begin{equation}
\begin{aligned}
\begin{bmatrix}
				\mathbf{A}(t)-\lambda(t) \mathbf{B}(t) & -\mathbf{B}(t)\mathbf{v}({t}) \\
				\mathbf{c}^{\mathrm{H}} \mathbf{B}(t) & 0
\end{bmatrix}
\begin{bmatrix}
				\mathbf{v}'(t)\\\lambda'(t)
\end{bmatrix}
				=\\
\begin{bmatrix}
				-\mathbf{A}'(t)\mathbf{v}(t)\! + \!\lambda(t) \mathbf{B}'(t) \mathbf{v}(t)\\ - \mathbf{c}^{\mathrm{H}} \mathbf{B}'(t) \mathbf{v}(t),
\end{bmatrix},
\end{aligned}
\label{eq:derEP}
\end{equation}
we obtain the derivatives of the eigenpair with respect to~$t$, i.e. $\mathbf{v}'(t)$ and~$\lambda'(t)$, when provided the derivatives of the system matrices.
If these are not given analytically, they can be approximated, e.g., via finite differences as employed in~\cite{Ziegler_2023aa}, or, for the case of shape deformations, can be computed using shape derivatives~\cite{Ziegler_2023ab}.
Then, using the derivatives $\mathbf{v}'(t)$ and~$\lambda'(t)$ we estimate the eigenpair at the next parameter $t+h$ with suitable step size~$h$ by first-order Taylor expansion.
Subsequently, we solve~\eqref{eq:eigproblem} for $t+h$ and match the solutions with the estimated eigenvector by comparison based on a correlation factor~\cite{Jorkowski_2018aa}.

Note, that depending on the step size $h$ and the number of degrees of freedom, this procedure is computationally expensive.
Therefore, we propose to perform the eigenvalue tracking on a reduced model.

\section{Reduced Basis Approximation}\label{sec:rb}
\begin{figure*}[thbp]
\centering
\includegraphics{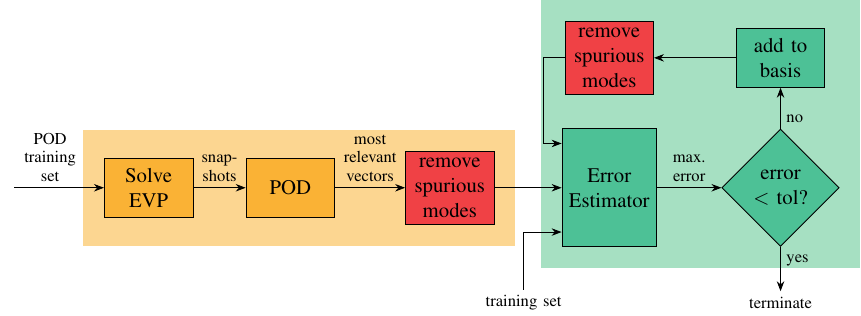}
\caption{Two-phase algorithm to construct the reduced basis with additional steps to remove the spurious modes in both phases.}
\label{fig:flowchart}
\end{figure*}

The reduced basis method approximates the full space of size~$\mathcal{N}$ by a reduced basis of size~$N\ll\mathcal{N}$ \cite{Patera_2007aa, Rozza_2008}.
The reduced basis must contain information such that the $K$ eigenvalues of interest are approximated sufficiently well at all parameter values~$t$.
Furthermore, since some eigenmodes appear in different polarizations, the algorithm is required to consider multiple eigenvalues. 
These requirements are met by the algorithm by Horger et al.~\cite{Horger_2017aa}. However, since it is restricted to fixed computational domains and demonstrated on a linear elasticity problem, we propose the extension to Maxwell eigenvalue problems and parameter-dependent domains.
A combination of the reduced basis approach for elliptic partial differential equations with a tracking procedure and a sparse grid-based adaptive sampling is proposed in~\cite{Alghamdi}.
Horger's algorithm is structured in two phases that implement different approaches, which we will describe in the following.
Our extended algorithm is illustrated in \autoref{fig:flowchart}.

\subsection{Initialization with Proper Orthogonal Decomposition}
For the initialization, we build a first basis using Proper Orthogonal Decomposition (POD), following e.g.~\cite{Kahlbacher_2007}.
The idea is based on a Singular Value Decomposition (SVD), where we orthonormalize the solutions and choose the most relevant functions for the initial basis.

We use the method of snapshots, i.e. we choose a sufficiently large training set~$\Xi_{\mathrm{train}}^{\mathrm{POD}}
=\{t_1,\ldots,t_{\Npod}\}$ at which to evaluate the system matrices and solve the parameterized eigenvalue problem~\eqref{eq:eigproblem} for each parameter $t_i\in\Xi_{\mathrm{train}}^{\mathrm{POD}}$.
We then gather the first~$K$ solution eigenvectors at each parameter in~$\Xi_{\mathrm{train}}^{\mathrm{POD}}$, also called snapshots that we now denote by~$\mathbf{y}$, in a matrix $\mathbf{Y} \coloneqq \left[\mathbf{y}_1, \ldots, \mathbf{y}_\Npod\right]$.
Using the symmetric, positive-definite mass matrix $\mathbf{B}$, which induces a weighted inner product, we compute
\begin{equation}
    \mathbf{K} = \mathbf{Y}^\top \mathbf{B} \mathbf{Y}
\end{equation}
and then calculate the eigenvalue decomposition of $\mathbf{K}$.
In this way, we obtain the eigenvectors $\bar{\mathbf{u}}_i$ and the eigenvalues~$\bar{\lambda}_i$.
Note, that we sort the eigenvalues $\bar{\lambda}_i$ such that $\bar{\lambda}_1 > \bar{\lambda}_2> \ldots> \bar{\lambda}_\Npod$.
We then compute $\mathbf{z}_i = \mathbf{Y}\bar{\mathbf{u}}_i/\sqrt{\bar{\lambda}_i}$ which yields the orthonormalized POD basis $\{\mathbf{z}_i\}_{i=1}^{\Npod}$ and we choose the first $\Ninit$ basis vectors (i.e. the ones corresponding to the $\Ninit$ largest eigenvalues).
Hence, we obtain the initial reduced basis $V_\Ninit \coloneqq \mathrm{span} \{\mathbf{z}_1, \ldots, \mathbf{z}_\Ninit\}$ and assemble the basis vectors in a matrix $\mathbf{Z} \coloneqq ({\mathbf{z}_1},\ldots, \mathbf{z}_{\Ninit})$.
Finally, we obtain the reduced stiffness and mass matrices
\begin{equation}
\mathbf{A}_{\mathrm{red}}(t) \coloneqq \mathbf{Z}^\top \mathbf{A}(t)\mathbf{Z},
\quad
\mathbf{B}_{\mathrm{red}}(t) \coloneqq \mathbf{Z}^\top \mathbf{B}(t)\mathbf{Z}.
\end{equation}
Note, that we can use the matrix~$\mathbf{Z}$ to upscale the reduced eigenvector $\mathbf{v}_{\mathrm{red}}(t)$ obtained by solving the reduced eigenvalue problem on $\mathbf{A}_{\mathrm{red}}(t)$ and $\mathbf{B}_{\mathrm{red}}(t)$ to the original space by the matrix-vector product 
\begin{equation}
\mathbf{v}(t) \approx \mathbf{Z} \mathbf{v}_{\mathrm{red}}(t).    
\label{eq:upscaleV}
\end{equation}

\subsection{Greedy Strategy for the Reduced Basis}
After the initialization via POD, the reduced basis is extended based on a greedy strategy. 
Iteratively, the (orthonormalized) eigenvector is added to the basis, which is approximated with the least accuracy according to an a-posteriori error estimator.
The formulation of the appropriate error estimator is one of the main features of the algorithm by Horger et al.~\cite{Horger_2017aa}. 

\subsubsection{The Error Estimator} 
The error estimator developed in~\cite{Horger_2017aa} takes several factors into account and is evaluated for each eigenvalue~$\lambda_{\mathrm{red},i}$ computed by the reduced system with $i = 1, \ldots, K$.
It is designed in order to also account for multiple eigenvalues.
Besides the eigenvalue~$\lambda_{\mathrm{red},i}$ itself, we consider the relative distance 
\begin{equation}
d_i(t) \coloneqq \left\lvert\frac{\lambda_{\mathrm{red},l}(t) - \lambda_{\mathrm{red},i}(t)}{\lambda_{\mathrm{red},l}(t)}\right\rvert
\end{equation}
to the closest neighboring eigenvalue $\lambda_{\mathrm{red},l}(t)$ in the sense of~\cite[(3.10)]{Horger_2017aa}, which approximates a different eigenvalue in the high-fidelity system than $\lambda_{\mathrm{red},i}(t)$. 
I.e.,~$d_i(t)$ indicates the smallest distance to an eigenvalue~$\lambda_{\mathrm{red},l}(t)$ that is not a multiple one of~$\lambda_{\mathrm{red},i}(t)$.
Lastly, we compute the residual~$\mathbf{r}_i$ of the approximation error in the high-fidelity space.
Note, that although the error estimator must be computed for each eigenvalue of interest at each parameter $t$ in the training set, this can be done efficiently since the analysis can be performed on the reduced space by computing
\begin{equation}
\begin{aligned}
    \mathbf{r}_{i}(t) &= \mathbf{A}(t)\left( \sum_{n=1}^N (\mathbf{v}_{\mathrm{red},i}(t))_n \mathbf{z}_n   \right) \\
    &- \lambda_{\mathrm{red},i}(t) \mathbf{B}(t) \left( \sum_{n=1}^N (\mathbf{v}_{\mathrm{red},i}(t))_n \mathbf{z}_n   \right),
\end{aligned}
\end{equation}
where $N$ indicates the current size of the basis. 
Then, the error estimator reads
\begin{equation}
    \eta_i(t) \coloneqq \frac{\mathbf{r}_i(t)^\top \mathbf{B}(t)\mathbf{r}_i(t)}{g(t)d_i(t) \lambda_{\mathrm{red},i}(t)},
\label{eq:error}
\end{equation}
{where we choose the parameter-dependent coercivity constant $g(t)=1$. 
We refer the reader to~\cite[Sec. 3 and 5]{Horger_2017aa} for a detailed description of the error estimator in a more general formulation and of the parameter-dependent coercivity constant, respectively.

\subsubsection{The Greedy Algorithm}
In the following, we discuss the greedy strategy to build the reduced basis, suitable to track the $K$ smallest eigenvalues of interest simultaneously.
Since the \mbox{$K$-th} eigenvalue might be a multiple eigenvalue, we need to include~$K+\tau$ eigenvalues in our considerations, where~$\tau$ denotes the maximum multiplicity of an eigenvalue.
Therefore, we initialize the algorithm using a POD with a sufficient size $\Ninit$.
In their work, Horger et al. suggest choosing at least $\Ninit \geq 1.5\cdot (K + \tau)$ to ensure a sufficient reliability of the error estimator.
This initial basis is now extended using the greedy strategy.

We select a training set~$\Xi_{\mathrm{train}}$ of $\Ntrain$ parameters and then for each instance in $\Xi_{\mathrm{train}}$, we evaluate the error estimate $\eta_i(t)$ for each $i =1, \ldots, K$.
The basic idea of the greedy strategy is to add the eigenspace of the parameter, for which the error estimate~\eqref{eq:error} is maximal.
If the corresponding eigenvalue has a multiplicity greater than one, all eigenvectors of the eigenspace are added to the basis.
The algorithm terminates when all~$K$ smallest eigenvalues of interest are approximated with the desired accuracy, i.e. when the maximum value of the error estimator is smaller than a predefined error tolerance.

\section{Removal of Spurious Modes}\label{sec:spuriousModes}
Albeit discretizing the solution with Nédélec edge elements, the projection to the reduced space is not exactly divergence-free, i.e. we add potentially spurious modes to the basis.
These inaccuracies are caused by numerical errors from various approximations, such as truncation errors, which are adversely amplified in the orthonormalization in the POD.
We observe these spurious modes as an added gradient field to the solutions, making them no longer divergence-free.
Hence, we add an additional step into the algorithm to remove these spurious solutions after the initialization of the basis with POD and in each iteration of the greedy strategy removing the spurious solutions from the basis.
In the following, we suggest the three options~\ref{sec:gramschmidt}-\ref{sec:treecotree}. 
An alternative route is using the mixed formulation of Kikuchi~\cite{Kikuchi_1987aa} and adding a supremizer stabilization as introduced in~\cite{Ballarin} for a Navier-Stokes setting.

\subsection{Orthogonalization} \label{sec:gramschmidt}
The first option is based on the idea to orthogonalize the snapshots with respect to the gradient space and hence to remove gradient fields from the basis.
Let us recall that the basis~$\mathbf{Z}$ obtained from the POD is an orthonormal basis due to the construction using an eigenvalue solver and suitable normalization. 
However, the basis~$\mathbf{Z}$ may have lost the orthogonality with respect to the gradient space which we now need to restore.
To this end, let $\mathbf{G}$ denote the discrete representation of the gradient operator.
We then orthogonalize each snapshot with respect to~$\mathbf{G}$, e.g. with a modified Gram-Schmidt iteration
\begin{equation}
    \mathbf{z}_{\mathrm{orth},i}  = \mathbf{z}_i - \sum_{j=1}\left(\mathbf{G}_{:,j}^\top \mathbf{B}(t)\mathbf{z}_i \right) \mathbf{G}_{:,j},
    \label{eq:gramschmidt}
\end{equation}
where $\mathbf{G}_{:,j}$ denotes the $j$-th column of $\mathbf{G}$.
We iterate over each column $j$ of $\mathbf{G}$ and apply \eqref{eq:gramschmidt} individually to each vector~$\mathbf{z}_i$ in the reduced basis.
When we gather each vector in the resulting matrix~$\mathbf{Z}_{\mathrm{orth}}$, this basis will be divergence-free, without spurious modes.
Nonetheless, note that also the modified version of the Gram-Schmidt algorithm is still susceptible to numerical instabilities. 
In~\eqref{eq:gramschmidt}, we choose a fixed value $t=0$ for the orthogonalization, which ties the divergence freeness to the domain $\Omega(0)$. 
This choice results in insufficient accuracy of the computed eigenvalues at $t=1$ as can be seen in \autoref{fig:tracking}. 
Here, the orange lines indicating the eigenvalues computed on the reduced system notably deviate from the black lines indicating the high-fidelity solutions.
On the other hand, continuously changing $t$ would lead to a parameter-dependent reduced basis, which is computationally unattractive.

\subsection{Projection}
Alternatively, a grad-div-type projection operator can be employed to remove the non-trivial divergence in the solutions as suggested in~\cite{Clemens_2005aa}.
To this end, we introduce the matrix~$\mathbf{C}$ with 
\begin{equation}
    \mathbf{C}_{i,j}(t) = \int_{\Omega(t)} \varepsilon  \, \textrm{grad}\, p_j \cdot  \mathbf{w}_i \, \mathrm{d}\mathbf{x}, 
\end{equation}
where the basis functions~$p_j$ are chosen from a finite-dimensional subspace of $H_0^1(\Omega(t))$.
Note, that this corresponds to the off-diagonal block matrices of the mixed problem formulation~\cite{Kikuchi_1987aa}.
We remark that, furthermore, the relation $\mathbf{C}^\top(t) = \mathbf{B}(t)\mathbf{G}$ holds.

Then, the projection operator reads
\begin{equation}
    \mathbf{P} \coloneqq \mathbf{I} - \mathbf{G}\left(\mathbf{C}(t)\mathbf{G}\right)^{-1}\mathbf{C}(t)
\end{equation}
and applied to the basis~$\mathbf{Z}$, fraught with spurious gradients, gives the divergence-free basis vectors
\begin{equation}
    \mathbf{z}_{\mathrm{orth},i} = \mathbf{Pz}_{i}.
\end{equation}
By applying this projector to the reduced basis vectors, we note improved orthogonality of the basis vectors with respect to the discrete gradient matrix~$\mathbf{G}$.
Nevertheless, the accuracy of the eigenvalues of the reduced system still depends on the chosen parameter value for $t$ at which to evaluate $\mathbf{C}(t)$.
Therefore, also the projector results in insufficient accuracy of the eigenvalues obtained by the reduced system for a fixed value of $t$ and shows a similar behavior as variant~\ref{sec:gramschmidt} in the tracking procedure.

\subsection{Tree-Cotree Decomposition}\label{sec:treecotree}
In order to remove spurious modes which appear as an added gradient field, also a tree-cotree gauge as proposed in~\cite{Manges_1995aa} can be employed.
The tree-cotree gauge relies on finding a spanning tree and the cotree on the edges of
a mesh to decompose the degrees of freedom, which, when using Nédélec-type basis functions, are associated with the edges of the mesh.
Hence, it only depends on the topology of the mesh and is independent of a mesh deformation, when avoiding remeshing, which is naturally given by Isogeometric Analysis.
We condense the system to the dimension of the degrees of freedom corresponding to the cotree, which results in removing the gradient-field, i.e. the non-trivial nullspace.
To this end, we denote the set of indices belonging to the tree and cotree edges with~$T$ and~$C$, respectively, and define the matrix
\begin{equation}
    \mathbf{H} \coloneqq [\mathbf{A}_{CC}, \mathbf{A}_{CT}],
\end{equation}
where we only include the rows of the stiffness matrix corresponding to the degrees of freedom of the cotree edges.
By performing a change of variables via 
\begin{equation}
    \mathbf{v} = \mathbf{B}^{-1}\mathbf{H}^\top\mathbf{y},
    \label{eq:variableChange}
\end{equation}
 we obtain the new system
\begin{equation}
    \hat{\mathbf{A}}\mathbf{y} = \lambda \hat{\mathbf{B}}\mathbf{y},
    \label{eq:cotreeSystem}
\end{equation}
where
\begin{align}
    \hat{\mathbf{A}} &= (\mathbf{HB}^{-1})\mathbf{A}(\mathbf{HB}^{-1})^\top\\
    \hat{\mathbf{B}} &= (\mathbf{HB}^{-1})\mathbf{B}(\mathbf{HB}^{-1})^\top.
\end{align}
Then, by solving~\eqref{eq:cotreeSystem}, we exclude spurious modes.
Using~\eqref{eq:variableChange}, we transform back to the original space.
Since this procedure is automatically valid for all domains $\Omega(t)$, the tree-cotree gauge therefore achieves the highest accuracy and in the resulting plot, the tracking on the reduced system is not distinguishable from the high-fidelity solutions. 

\section{Numerical Results}\label{sec:results}
We demonstrate our algorithm for the 1-cell TESLA cavity using the algebraic tracking~\cite{Ziegler_2023aa}.
We build a reduced basis to approximate the first ten eigenvalues along the shape deformation to the pillbox cavity. 
All implementations are done in MATLAB\textsuperscript{\textregistered} using the GeoPDEs package~\cite{Vazquez_2016}.
The full discretized model has 1468 degrees of freedom (when discretized with curl-conforming splines of degree $2$). 
For the scalar deformation parameter $t$, we choose an equidistant training set $\Xi_{\mathrm{train}}$ of size 100. 
For the test set~$\Xi_{\mathrm{test}}$ we choose 200 random values. 
We add $50$ initial vectors to the basis, then perform the greedy algorithm to add further eigenvectors to the basis and observe the average error 
\begin{equation}
\mathcal{E}_{i,\mathrm{av}}= \frac{1}{\#\Xi_{\mathrm{test}}} \sum_{t \in \Xi_{\mathrm{test}}} \frac{\lambda_{\mathrm{red}}(t)-\lambda_i(t)}{\lambda_i(t)} 
\end{equation}
for the eigenvalues~$i = 1,\ldots, K$, of the test set, c.f. \autoref{fig:errors}, where $K=10$. 
It can be seen that after adding approximately 150 eigenvectors, we reach the level of accuracy of the original system and achieve no further improvement by increasing the basis.

When now considering $K=5$ for the tracking from the one TESLA-cell to the pillbox cavity, also the optimal size of the reduced basis decreases as observed from our investigations in \autoref{fig:errors_K5}. 
Here, we note that the average error for the eigenvalues of interest does not improve further after adding $80$ vectors to the basis.
Therefore, we apply the reduced basis of dimension $80$ in the tracking of the first five eigenmodes along the shape deformation and remove the spurious eigenmodes via the tree-cotree gauge, see \autoref{fig:tracking}.
Thereby, we reduce the duration of the tracking (excluding the system matrix assembly and construction of the reduced basis, respectively) from \SI{14.7157}{\s} to \SI{0.2853}{\s}, which corresponds to a speed-up of $51.6$.
Further results can be found in \autoref{tab:Results}, where we compare the computational time of solving one eigenvalue problem and performing the tracking on the high-fidelity system, the high-fidelity system on the cotree edges, the reduced system gauged with the tree-cotree approach and the reduced system with orthogonalization using the modified Gram-Schmidt algorithm, respectively. 
All timings were measured on a workstation with an Intel\textsuperscript{\textregistered} CPU i7-3820 3.6-GHz processor and 13-GB RAM and averaged over $10$ runs.

\begin{table}[htbp]
    \centering
    \caption{Comparison of the computation times and speedups between the high-fidelity systems with and without a gauge, the reduced system gauged with the tree-cotree gauge and the reduced system with orthogonalization using the modified Gram-Schmidt algorithm, respectively.}
\begin{tabular}{l|c|c||c|c|}
    & \mbox{High-fidelity} & \mbox{High-fidelity} & RB with & RB with\\ 
     & w.o. gauge & on Cotree & Tree-Cotree &  Gram-Schm.\\ \hline 
 \#DoF  & $2148$ & $1468$ & $80$  &  $80$ \\ \hline
 EVP & \SI{0.1132}{\s} & \SI{0.5294}{\s} & \SI{0.0017}{\s}  &  \SI{0.0044}{\s}    \\
 Speedup &  1 & $0.21$ & $66.6$  &   $25.7$   \\ \hline
 Tracking &  \SI{14.7157}{\s}  & \SI{19.4983}{\s} & \SI{0.2853}{\s}  & \SI{0.3852}{\s}     \\ %
 Speedup&  1  & $0.7547$ & $51.6$ &  $38.2$    
\end{tabular}
    
    \label{tab:Results}
\end{table}

We first compare the performance of solving the high-fidelity systems, i.e., \eqref{eq:eigproblem} and the cotree-condensed system~\eqref{eq:cotreeSystem}. We note that although the number of degrees of freedom is significantly smaller, solving only for the cotree system is, as expected, less efficient due to the denser matrices. However, comparing the tree-cotree gauge and the orthogonalization via Gram-Schmidt on the reduced system, we observe that the tree-cotree-gauged system performs significantly better than the system where we removed the spurious modes by orthogonalization.

\begin{figure}
\includegraphics[]{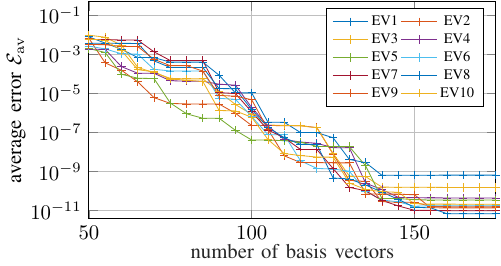}
\caption{Average error of the first ten approximated eigenvalues in the 1-cell TESLA cavity with 50 initial basis vectors.}
\label{fig:errors}
\end{figure}

\begin{figure}
\includegraphics[]{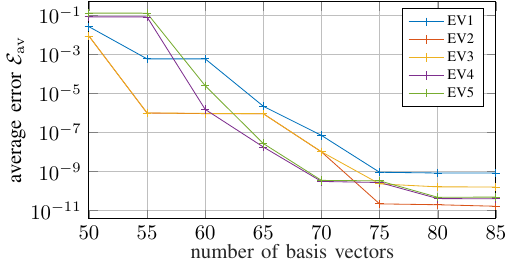}
\caption{Average error of the first five approximated eigenvalues in the 1-cell TESLA cavity with 50 initial basis vectors.}
\label{fig:errors_K5}
\end{figure}

\section{Conclusion}\label{sec:conclusion}
In this paper, we have extended a state-of-the-art reduced basis algorithm for parameter-dependent domains and the Maxwell eigenvalue problem.
We have demonstrated that in this setting, we encounter spurious eigenmodes after the projection to the reduced space and we proposed approaches to remove the spurious solutions, of which we found the tree-cotree gauge to be the most accurate and time-efficient. 
We exemplified our algorithm in the automatic classification of eigenmodes based on eigenvalue tracking and shape morphing. 
Especially when refining the model in order to resolve the more detailed structures of the higher modes or when the step size of the tracking algorithm is decreased, the computations may get very expensive.
In these cases, tracking on the reduced basis is particularly beneficial.
Thereby, the reduced basis approach also permits real-time and many-query applications of parameterized partial differential equations, which allows for, e.g., more efficient design optimization of a device.

\section*{Acknowledgment}
We thank Markus Clemens and Carlo De Falco for the many fruitful discussions.
We thank Melina Merkel for the support, the many fruitful discussions as well as for proofreading the manuscript.

This work is supported by the Graduate School CE within the Centre for Computational Engineering at TU Darmstadt. 

\bibliographystyle{IEEEtran}

\begin{thebibliography}{10}

\bibitem{Lui_1997aa}
S.~H. Lui, H.~B. Keller, and T.~W.~C. Kwok, ``Homotopy method for the large,
  sparse, real nonsymmetric eigenvalue problem,'' \emph{{SIAM} J. Matrix. Anal.
  Appl.}, vol.~18, no.~2, pp. 312--333, Apr. 1997.

\bibitem{Jorkowski_2018aa}
P.~Jorkowski and R.~Schuhmann, ``Mode tracking for parametrized eigenvalue
  problems in comp. electromagn.'' in \emph{2018 International Applied
  Computational Electromagnetics Society ({ACES}) Symposium}, B.~M. Notaros,
  Ed.\hskip 1em plus 0.5em minus 0.4em\relax IEEE, Mar. 2018, p. 17803096.

\bibitem{Aune_2000aa}
B.~Aune, R.~Bandelmann, D.~Bloess, B.~Bonin, A.~Bosotti, M.~Champion,
  C.~Crawford, G.~Deppe, B.~Dwersteg, D.~A. Edwards, H.~T. Edwards,
  M.~Ferrario, M.~Fouaidy, P.-D. Gall, A.~Gamp, A.~Gössel, J.~Graber,
  D.~Hubert, M.~Hüning, M.~Juillard, T.~Junquera, H.~Kaiser, G.~Kreps,
  M.~Kuchnir, R.~Lange, M.~Leenen, M.~Liepe, L.~Lilje, A.~Matheisen, W.-D.
  Möller, A.~Mosnier, H.~Padamsee, C.~Pagani, M.~Pekeler, H.-B. Peters,
  O.~Peters, D.~Proch, K.~Rehlich, D.~Reschke, H.~Safa, T.~Schilcher,
  P.~Schmüser, J.~Sekutowicz, S.~Simrock, W.~Singer, M.~Tigner, D.~Trines,
  K.~Twarowski, G.~Weichert, J.~Weisend, J.~Wojtkiewicz, S.~Wolff, and
  K.~Zapfe, ``Superconducting {TESLA} cavities,'' \emph{Phys. Rev. Accel.
  Beams}, vol.~3, no.~9, p. 092001, 2000.

\bibitem{Ziegler_2023aa}
A.~Ziegler, N.~Georg, W.~Ackermann, and S.~Schöps, ``Mode recognition by shape
  morphing for {Maxwell}'s eigenvalue problem in cavities,'' \emph{{IEEE}
  Trans. Antenn. Propag.}, vol.~71, no.~5, pp. 4315--4325, 2023.

\bibitem{Horger_2017aa}
T.~Horger, B.~Wohlmuth, and T.~Dickopf, ``Simultaneous reduced basis
  approximation of parameterized elliptic eigenvalue problems,'' \emph{ESAIM:
  Mathematical Modelling and Numerical Analysis}, vol.~51, no.~2, pp. 443--465,
  2017.

\bibitem{Vazquez_2010aa}
R.~Vázquez and A.~Buffa, ``Isogeometric analysis for electromagnetic
  problems,'' \emph{{IEEE} Trans. Magn.}, vol.~46, no.~8, pp. 3305--3308, 2010.

\bibitem{Ziegler_2023ab}
A.~Ziegler, M.~Merkel, P.~Gangl, and S.~Schöps, ``On the computation of
  analytic sensitivities of eigenpairs in isogeometric analysis,''
  \emph{Computer Methods in Applied Mechanics and Engineering}, vol. 409, p.
  115961, 2023. [Online]. Available:
  \url{https://www.sciencedirect.com/science/article/pii/S0045782523000841}

\bibitem{Patera_2007aa}
A.~T. Patera and G.~Rozza, \emph{Reduced Basis Approximation and a Posteriori
  Error Estimation for Parametrized PDEs}.\hskip 1em plus 0.5em minus
  0.4em\relax MIT, 2007. [Online]. Available:
  \url{http://augustine.mit.edu/methodology/methodology_book.htm}

\bibitem{Rozza_2008}
G.~Rozza, D.~B.~P. Huynh, and A.~T. Patera, ``Reduced basis approximation and a
  posteriori error estimation for affinely parametrized elliptic coercive
  partial differential equations,'' \emph{{Archives of Computational Methods in
  Engineering}}, vol.~15, no.~3, pp. 229 -- 275, Sep. 2008. [Online].
  Available: \url{https://hal.science/hal-01722593}

\bibitem{Alghamdi}
M.~M. {Alghamdi}, D.~{Boffi}, and F.~{Bonizzoni}, ``{A greedy MOR method for
  the tracking of eigensolutions to parametric elliptic PDEs},'' \emph{arXiv
  e-prints}, p. arXiv:2208.14054, Aug. 2022.

\bibitem{Kahlbacher_2007}
S.~V. Martin~Kahlbacher, ``{Galerkin proper orthogonal
  decomposition methods for parameter dependent elliptic systems},''
  \emph{{Discussiones Mathematicae, Differential
  Inclusions, Control and Optimization}}, vol.~27, no.~1, pp. 95--117, 2007.
  [Online]. Available: \url{http://eudml.org/doc/271156}

\bibitem{Kikuchi_1987aa}
F.~Kikuchi, ``Mixed and penalty formulations for finite element analysis of an
  eigenvalue problem in electromagnetism,'' \emph{Computer Methods in Applied
  Mechanics and Engineering}, vol.~64, pp. 509--521, 10 1987.

\bibitem{Ballarin}
F.~Ballarin, A.~Manzoni, A.~Quarteroni, and G.~Rozza, ``Supremizer
  stabilization of {POD}–{Galerkin} approximation of parametrized steady
  incompressible {Navier}–{Stokes} equations,'' \emph{International Journal
  for Numerical Methods in Engineering}, vol. 102, no.~5, pp. 1136--1161, 2015.
  [Online]. Available:
  \url{https://onlinelibrary.wiley.com/doi/abs/10.1002/nme.4772}

\bibitem{Clemens_2005aa}
M.~Clemens, ``{Large systems of equations in a
  discrete electromagnetism: formulations and numerical algorithms},''
  \emph{{IEE Proceedings - Science, Measurement and
  Technology}}, vol. 152, pp. 50--72(22), March 2005. [Online]. Available:
  \url{https://digital-library.theiet.org/content/journals/10.1049/ip-smt_20050849}

\bibitem{Manges_1995aa}
J.~Manges and Z.~Cendes, ``A generalized tree-cotree gauge for magnetic field
  computation,'' \emph{IEEE Transactions on Magnetics}, vol.~31, no.~3, pp.
  1342--1347, 1995.

\bibitem{Vazquez_2016}
R.~Vázquez, ``A new design for the implementation of isogeometric analysis in
  {Octave} and {Matlab}: {GeoPDEs} 3.0,'' \emph{Computers \& Mathematics with
  Applications}, vol.~72, no.~3, pp. 523--554, 2016. [Online]. Available:
  \url{https://www.sciencedirect.com/science/article/pii/S0898122116302681}

\end{thebibliography}

\end{document}